\documentclass[doublecol,A4paper]{epl2}

\title{Cooper pairs without `glue' in high-$T_c$ superconductors}
\shorttitle{Cooper pairs without `glue'}

\author{William Sacks\inst{1} \and Alain Mauger\inst{1,*} \and Yves Noat\inst{2}}
\shortauthor{W. Sacks, A. Mauger \& Y. Noat}

\institute{\inst{1} Institut de Min\'{e}ralogie, de Physique des
Mat\'{e}riaux, et de Cosmochimie (IMPMC), UMR 7590,

\inst{2} Institut des Nanosciences de Paris (INSP), UMR 7588,\\

Sorbonne Universit\'{e}s, UPMC Paris 6, \\
4 place Jussieu, 75252 Paris Cedex 05, France \\

\inst{*}Corresponding author\,: alain.mauger@impmc.jussieu.fr}


\abstract{We address the origin of the Cooper pairs in high-$T_c$
cuprates and the unique nature of the superconducting (SC)
condensate. Itinerant holes in an antiferromagnetic background form
pairs spontaneously, without any `glue', defining a new quantum
object the `pairon'. In the incoherent pseudogap phase, above $T_c$
or within the vortex core, the pairon binding energies are
distributed statistically, forming a `Cooper-pair glass'. Contrary
to conventional SC, it is the mutual pair-pair interaction that is
responsable for the condensation. We give a natural explanation for
the {\it ergodic rigidity} of the excitation gap, being uniquely
determined by the carrier concentration $p$ and $J$. The phase
diagram can be understood, without spin fluctuations, in terms of a
single energy scale $\sim J$, the exchange energy at the
metal-insulator transition.}

\begin{document}

\maketitle

\vskip 2mm

{\it Introduction}.

\vskip 2mm

A remarkable aspect of the BCS superconducting (SC) state is its
universality, explaining a large number of properties quite
independently of the detailed composition of the material
\cite{PR_BCS1957}. The conventional SC state emerges from a general
metallic state leading to a macroscopic wave function responsable
for zero resistivity, perfect diamagnetism and Josephson quantum
effects \cite{PhysLett_Josephson1962}. The microscopic picture is
the condensation of Cooper pairs \cite{PR_Cooper1956} wherein the
binding energy is the order parameter, vanishing at $T_c$, and whose
$T=0$ value is material independent, with a universal ratio \,:
2\,$\Delta/k_B \,T_c \simeq 3.52$. Although conventional SC is well
understood, the pairing mechanism due to phonon exchange
\cite{ProcSocLon_Frohlich1952} is still hidden in the ground state
energy gap. However it is revealed in the strong-coupling
quasiparticle (QP) excitation spectrum in the fine structure at the
phonon energies above the gap
\cite{PR_Giaever1962,PRL_McMillan1965}.

In marked contrast, high-$T_c$ superconductivity
\cite{ZPhys_Bednorz1986} emerges upon doping from an insulating
state dominated by antiferromagnetic (AF) interactions
\cite{ComRenPhys_Hirschfeld2015}. The macroscopic wavefunction
expresses the hallmark SC properties, but the microscopic mechanism,
in particular the pair formation, remains unknown. Moreover, the
relevant parameters are orders of magnitude different from the
conventional case\,: the nanometric coherence length, a large
spectral gap, large penetration length and, of course, very high
$T_c$. Strikingly, the magnitude of the spectral excitation gap, as
measured by tunneling \cite{PRL_renner1998_T,JPSJ_Sekine2016} and
photoelectron spectroscopies (\cite{NatPhys_Hashimoto2014} and refs.
therein), remains constant as a function of temperature up to and
just above $T_c$, in the pseudogap (PG) state (see
\cite{ReProgPhys_Timusk1999} and refs. therein), suggesting the
existence of pairing correlations above $T_c$
\cite{PRL_Randeria1992,Nat_Emery1995}.
 Thus, contrary to the conventional BCS scenario, the energy gap is clearly not the
order parameter.

In this work, we propose a simple mechanism for the pair formation
in high-$T_c$ cuprates.  We demonstrate that hole pairs in a
lightly-doped antiferromagnetic environnement are energetically
stable. Their binding energy is directly related to the exchange
energy $J$ of the surrounding electron spins. Thus, the hole pairing
arises spontaneously, without `glue', due to an effective quantum
potential well. This defines a new complex quantum state, a hole
pair coupled to its local AF environnement, the `pairon'.

\begin{figure}[t]
\centering \vbox to 9.5 cm{
\includegraphics[width=8 cm]{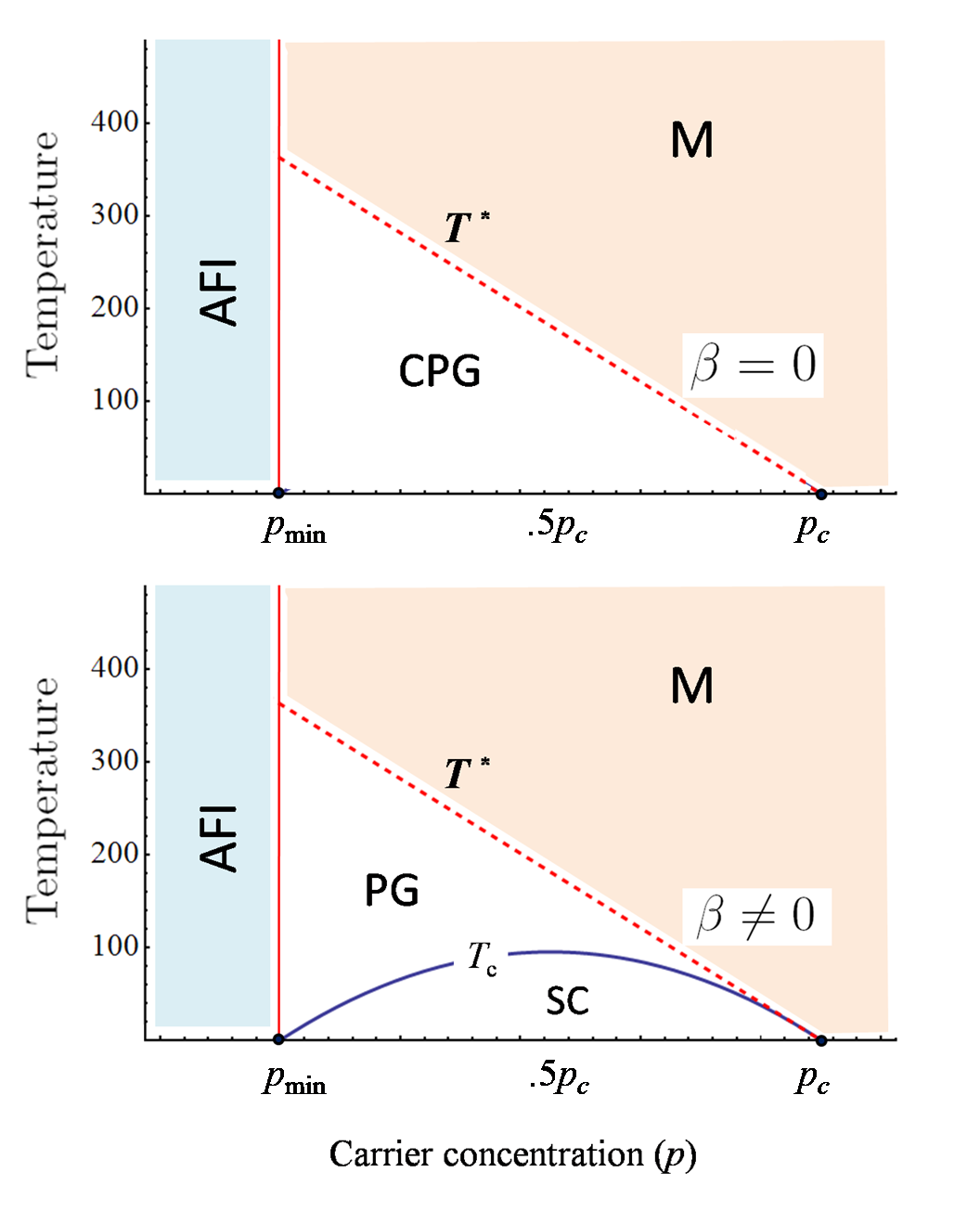}}
\caption{Phase diagram as a function of doping without pair-pair
interaction ($\beta=0$, upper panel) and with PPI ($\beta\neq 0$,
lower panel). As a result of the PPI the critical dome separates the
PG phase from the SC phase at lower
temperature.}\label{phasediagram}
\end{figure}

Beyond the AF phase, the pairons are in a disordered incoherent
state, a `Cooper-pair glass' wherein their binding energies are
statistically distributed. This state is revealed above $T_c$ in the
pseudogap (PG) phase, and also within the vortex core once SC
coherence is lost \cite{PRL_renner1998_B}. Contrary to the BCS
theory, the SC coherent state is achieved due to the mutual
pair-pair interaction $\beta$ (PPI), as shown in the phase diagram,
Fig.\,1, lower panel.

A novel fundamental constraint emerges in this work\,: the
condensate wave function is such that the hole-pair/anferromagnetic
state (pairon) must be homogeneous. As a result, the spectral gap is
pinned to the specific value $\Delta_p$ which is remarkably constant
with regards to perturbations such as temperature, magnetic field or
disorder. We show that the latter gap value is uniquely determined
by the doping $p$ and the exchange energy $J$. This pairing
constraint suggests the novel concept of `ergodic rigidity'.

Contrary to conventional SC, the pairing energy $\Delta_p$ is not
the condensation energy. Rather, it is the mutual interaction
between pairs that allows for a Bose-type condensation onto the
ground state \cite{SciTech_Sacks2015,EPJB_Sacks2016}. Consequently,
$\beta(T)$ is proportional to the condensate density which {\it
decreases with temperature} due to the pairon excitations, instead
of quasiparticles, to finally vanish at $T_c$. The PPI has the
required properties of an order parameter and in this work we show
that it is a unique function of $p$ and $J$.

The SC state is thus built from two inter-dependent phenomena, the
pairon formation and their mutual interaction, both depending on a
single energy scale $J$.

\vskip 0.2cm {\it Origin of Cooper pairs in cuprates} \vskip 0.2cm

In high-$T_c$ cuprates, superconductivity emerges upon doping from
the antiferromagnetic state at half filling i.e. one electron per
copper site. As a function of doping, it starts at a minimal value
$p_{min}\approx 0.05$, corresponding to the extinction of the
long-range magnetic order ($T_{N\acute{e}el} \sim 0$). In spite of
the vanishing of the AF order parameter, antiferromagnetism is known
to exist at the local scale, as indicated by the paramagnon
resonance even inside the SC dome \cite{PhysicaC_Rossat1991}. A
large number of measurements have confirmed the small SC coherence
length of the Cooper pairs in these materials ($\xi_{pair}\sim
1-2\,nm$). Thus, provided that the magnetic coherence length
$\xi_{AF}$ is large compared to the lattice constant, the exchange
energy $J$ is logically the dominant energy scale.

\begin{figure}[h!]
\centering \vbox to 5.0 cm{
\includegraphics[width=6.0 cm]{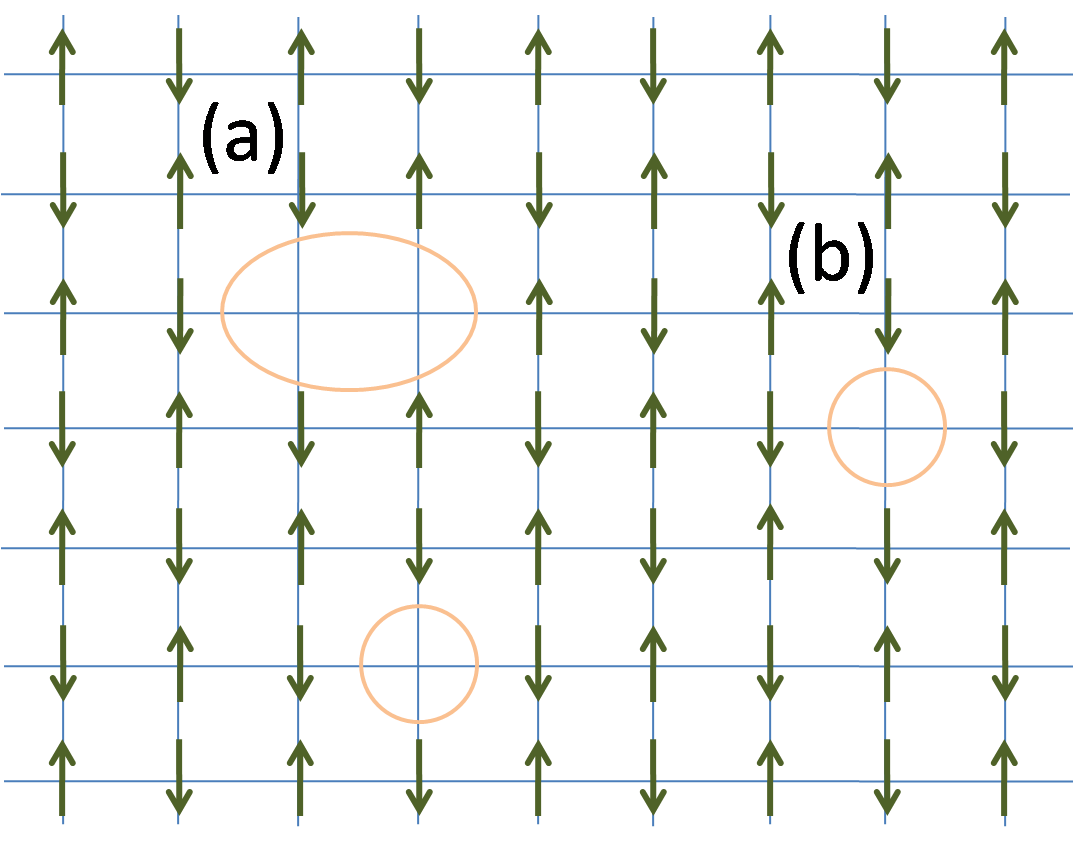}}
\caption{Representation of two separate holes as well as one hole
pair in an AF background. Due to the magnetic energy gain, the hole
pair configuration is favorable compared to two separate
holes.}\label{spins}
\end{figure}

In this AF background, pairs of holes are energetically more
favorable than individual holes (see Fig. \ref{spins}). Considering
only the magnetic potential energy for the square lattice,
$$H_{AF} = J\,\sum_{i\neq j}\vec{S_i}\cdot\vec{S_j}$$
where $i$ and $j$ are nearest neighbor occupied sites. Thus, two
individual holes have an energy 8\,$J$ with respect to the full AF
`sea', while a hole pair has an energy 6\,$J$, leading to a gain of
$J$ per hole. Thus, in the nearest-neighbor approximation, the
energy gain for a single hole pair in a perfect infinite AF
background is $J$ compared to two separate holes. This provides a
simple explanation for the pairing in these materials -- they arise
spontaneously without `glue', i.e. {\it without a boson exchange},
due to the surrounding AF background.

This effect can be viewed as a quantum potential well of depth $J$
for two holes. The question then arises whether 4 holes together, or
a more numerously populated island, would not be more stable. In
fact they are not since the interaction between pairs is repulsive.
Indeed, the effective potential well becomes rapidly nil when two
pairs are approached together. Given these considerations, the
expected stable solution for itinerant holes in an AF background
should be preformed pairs.

\begin{figure}[t]
\centering \vbox to 5.0 cm{
\includegraphics[width=7.0 cm]{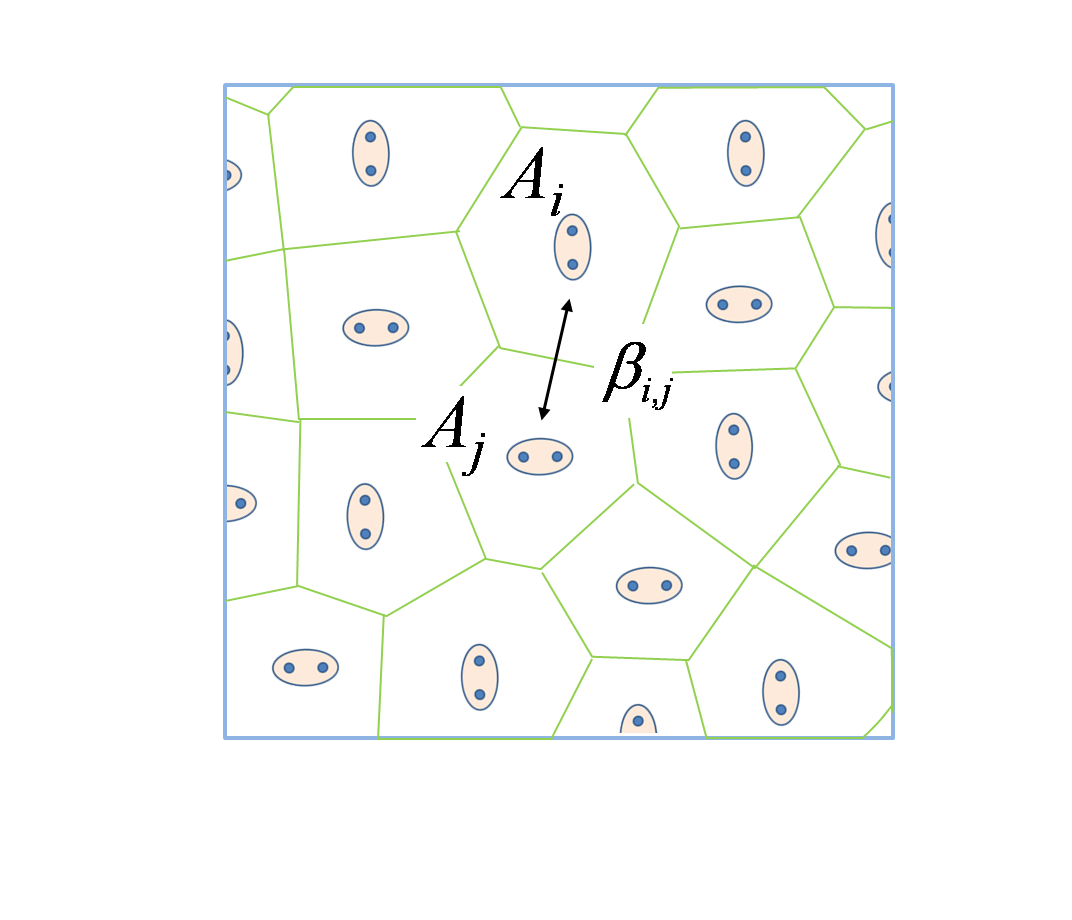}}
\caption{Snapshot spatial representation of the Cooper pair glass.
Each pair is associated with an AF area $\mathcal{A}_i$ which
determines the binding energy $\Delta_i$ of the
pairon.}\label{areas}
\end{figure}

We now consider a statistical ensemble of uncondensed pairs of
density $p/2$, each pair being surrounded by a number of nearest
neighbors as in Fig.\,\ref{areas}. Clearly, the binding energy for
each pair must be directly affected by the local AF environnement.
This conclusion is in good agreement with the experimental AF
correlation length as function of doping \cite{PRL_Ando2001}.
Remarkably, $\xi_{AF}\sim 1/\sqrt{p}$, which is also the typical
distance between pairs.

The simplest model, which works remarkably well, is that each
binding energy is uniquely determined by the area $\mathcal{A}_i$
available to each pair. This defines a new composite object, {\it
the pairon}, being a quantum state consisting of a hole pair
entangled with its local AF environnement. Thus, we take the $i$th
pairon binding energy $\Delta_i$ to be proportional to the
corresponding area $\mathcal{A}_i$ which, in a classical view, is
the Voronoi cell area in a two-dimensional distribution
\cite{JourMath_Voronoi1908}.

For large density, an additional consideration is necessary. As $p$
increases, and the characteristic areas decrease, there is a minimum
incompressible area $\mathcal{A}_c$ at which the local AF
environment is too small to create a potential well. In this limit,
the binding energy for a pairon vanishes. The minimum configuration
is one hole pair surrounded by 6 occupied copper sites, giving a
critical doping of $p_c = 2/8 \approx 0.25$. This agrees well with
the observed critical value for Bi$_2$Sr$_2$CaCu$_2$O$_{8+\delta}$ ($p_c \simeq 0.27$). This
effect is also related to the critical percolation threshold for
holes on a square lattice corresponding to 4 sites for each hole
\cite{JPhysChem_Tahir2010}.

For small densities, the critical point for SC onset is $p_{min}$
($p_{min} \simeq .05$ for Bi$_2$Sr$_2$CaCu$_2$O$_{8+\delta}$). This value corresponds to the
largest Voronoi cell area wherein the pairs can condense. At this
point, their binding energies are in fact maximum with a binding
energy close to $J$.

With these considerations, between these two limits, the pairing
energy for each pair $i$ is given by\,:
\begin{equation}
\Delta_i=J_{eff}\,(\mathcal{A}_i-\mathcal{A}_c)/<\mathcal{A}>
\label{equa_Jeff}
\end{equation}
where $J_{eff}$ is the effective exchange energy, $\mathcal{A}_i$ is
the Voronoi area for the $i$th pair (see Fig.\,\ref{areas}) and
$<\,>$ denotes the average. As a result, the mean binding energy
is\,:
\begin{equation}\label{delta0}
\Delta_0=<\Delta_i>=J_{eff}\times (1-\frac{p}{p_c})
\end{equation}
i.e. simply determined by the number of holes and $J_{eff}$. This
general relation shows that the dominant effect is the linear
decrease of the pair binding energy with the carrier concentration
to ultimately vanish at the critical point $p_c$, a universal
feature of high-$T_c$ superconductivity.

We now consider the many-body properties of the pairon system.

\vskip 0.2cm {\it The Cooper-pair glass state}
\vskip 0.2cm

Unconventional superconductivity in cuprates can be described by a
microscopic Hamiltonian of interacting preformed pairs
\cite{SciTech_Sacks2015}. In absence of pair-pair interactions, the
scenario is an incoherent state of preformed pairs that are
distributed in different energy states defined by their binding
energies $\Delta_i$. We call this state the Cooper-pair glass (CPG),
illustrated in Fig.\,\ref{areas}. It is quite analagous to the
Bragg-glass vortex phase \cite{PRB_Giamarchi1995,Nat_Klein2001},
unique to high-$T_c$ superconductivity, which is formed when the
vortex-vortex interaction (the PPI in the present case) is small
compared to any local disorder potential.

\begin{figure}[h!]
\centering \vbox to 5.0 cm{
\includegraphics[width=8. cm]{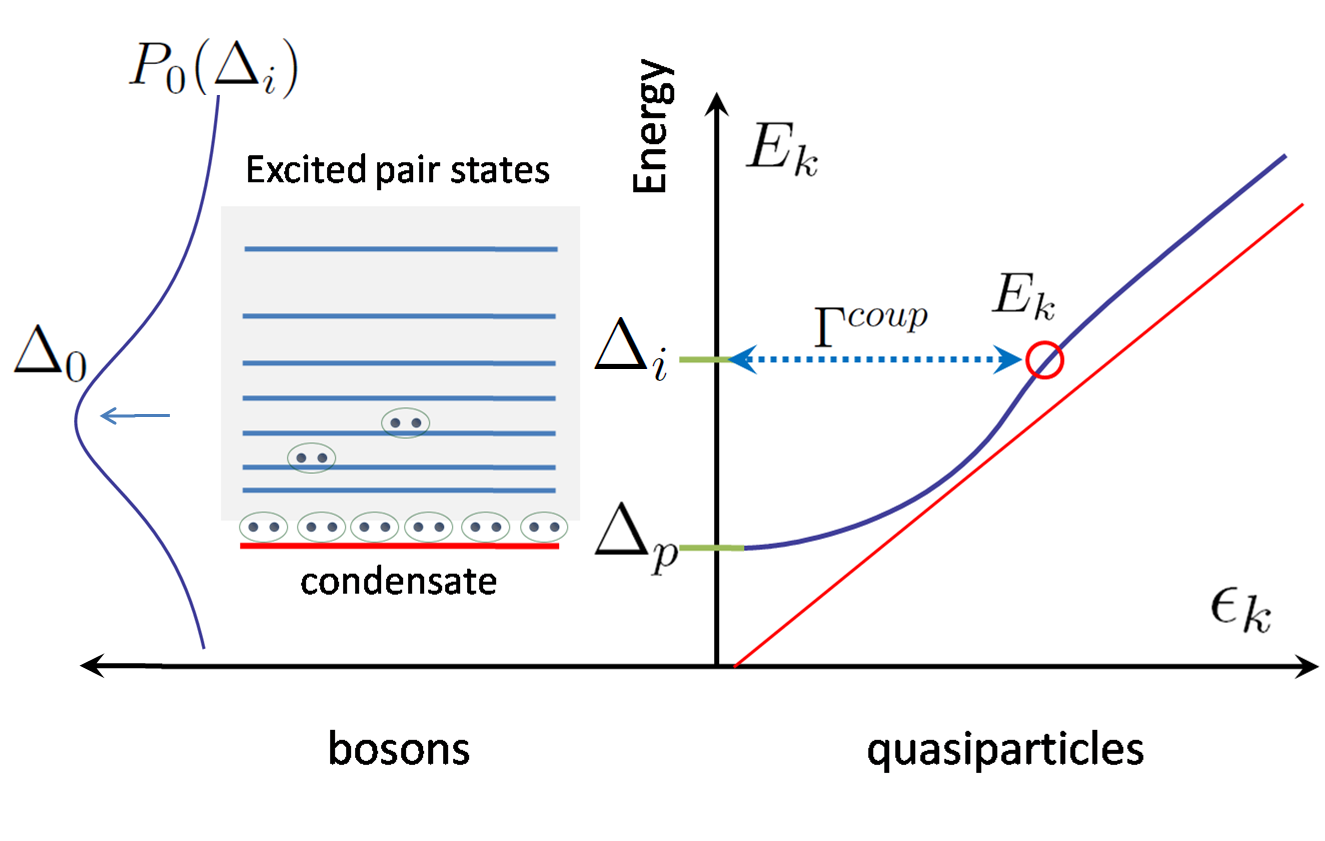}}
\caption{Boson/pair degrees of freedom (left side) and
fermion/quasiparticle excitations (right side). The degeneracy
between quasiparticle energy $E_k$ and the excited states energy
$\Delta_i$, leads to the dip singularity in the QP
DOS.}\label{quasiparticle}
\end{figure}

To understand the CPG state, characterized by $\Delta_i$, we note
that the CPG Hamiltonian can be formally mapped onto the
Suhl-Mathias-Walker model for multiband superconductors
\cite{PRL_Suhl1959} where $i$ is the band index. In our case, the
$i$th index refers to the given CPG pair state. The corresponding
density of states is very well described by a Lorentzian form\,:
\begin{equation}\label{P0}
P_0(\Delta_i) \propto \frac{\sigma_0^2}{(\Delta_i-\Delta_0)^2 +
\sigma_0^2}
\end{equation}
where $\Delta_0$ and 2\,$\sigma_0$ are respectively the median and
the width of the pair energy distribution (see
Fig.\,\ref{quasiparticle}, left panel). Again, $P_0(\Delta_i)$
reflects the variation of the possible configurations of the Voronoi
cells discussed previously.

\begin{table*}[t]
\centering
\begin{tabular}{@{}lccc@{}}
 SC parameters &\null & \null\ & \null\ \\
  \hline \hline \\
spectral gap& $\Delta_p$ &  \null & $J(1-p/p_c)$ \\
pair-pair int. & $1.2\,\beta^c$ & \null & $J(p/p_c)(1-p/p_c$) \\
dist. maximum & $\Delta_0$ & $\Delta_p + 1.2\,\beta^c$  & $J(1-p^2/{p_c}^2)$ \\
dist. width & $\sigma_0$ & $\Delta_0/2$ & $J(1-p^2/{p_c}^2)/2$ \\
dip energy & $E_{dip}$ & $\Delta_p + 2.4\,\beta^c$ & $J(1+2\,p/p_c)(1-p/p_c)$ \\
\end{tabular}
\caption{Universal dependence of the parameters with doping. All
parameters depend on the unique energy $J$. The doping $p$ and the
exchange energy $J$ are measured from the SC onset ($p_{min}$).
These equations are plotted in Fig.\,\ref{param}}
\label{table_param}
\end{table*}

\begin{figure}[h!]
\centering \vbox to 5.5 cm{
\includegraphics[width=8.5 cm]{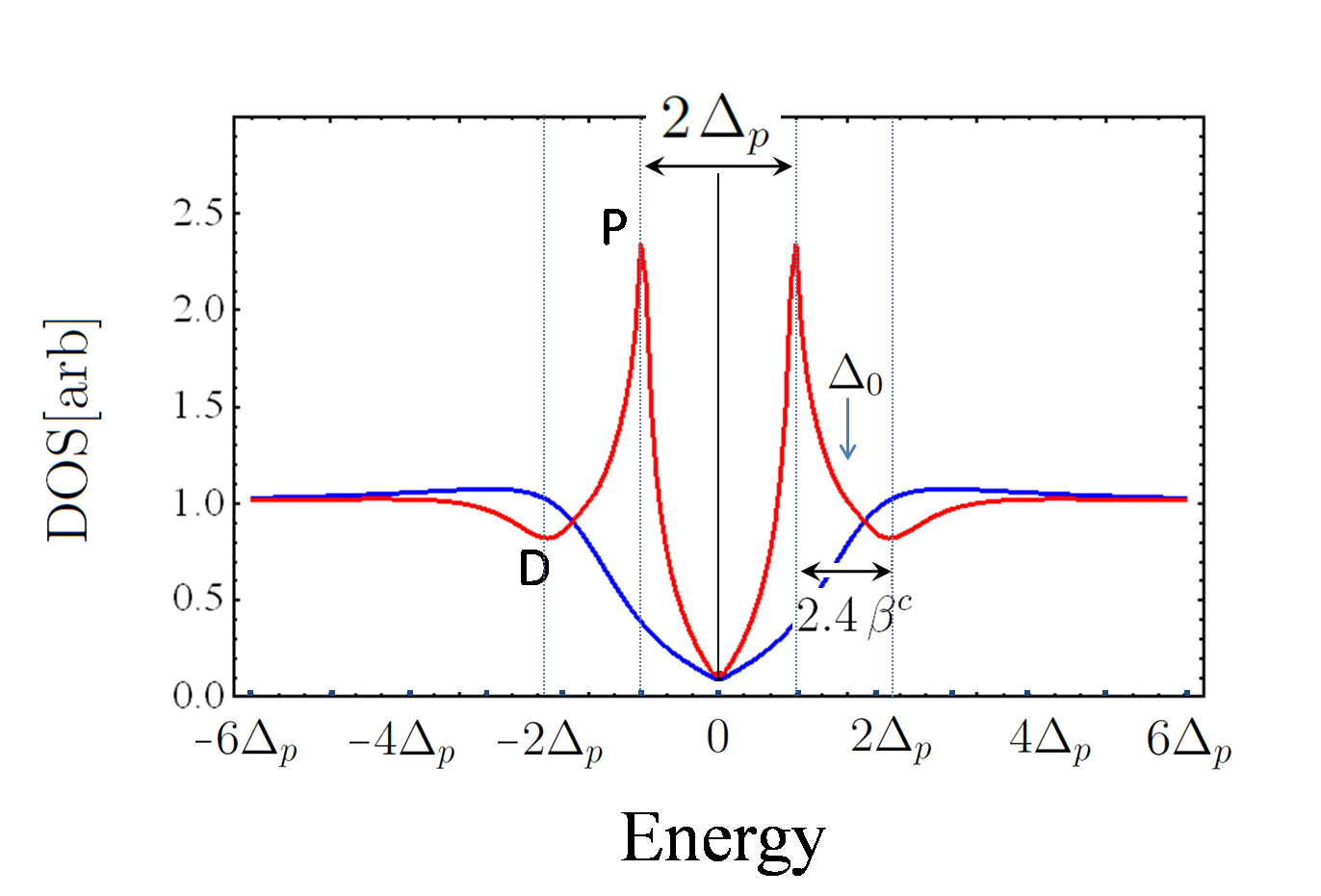}}
\caption{Quasiparticle excitation spectrum calculated within our
model in the Cooper glass state ($\beta^c=0$) and in the
superconducting state ($\beta^c\neq 0$) using equation
(\ref{gap_equa_QP}) near optimal doping. Note that the shape of the
coherence peak and dip structure in the SC state accurately fits the
experimental tunneling data
\cite{SciTech_Sacks2015}.}\label{spectre}
\end{figure}

Superconducting coherence arises as a result of the mutual
interactions between pairons, leading to a Bose-Einstein type
condensation into the single energy state $\Delta_i = \Delta_p$.
From the microscopic Hamiltonian \cite{SciTech_Sacks2015}, the
latter must satisfy the self-consistent gap equation :
\begin{equation}
\Delta_p = \Delta_0 - 2\,\beta^c\,P_0(\Delta_p)
\label{gap_equa}
\end{equation}
where $\beta^c$ is the interaction energy in the SC state described
below. In the absence of the PPI, i.e. $\beta^c = 0$, the pairons
are distributed in the energy states centered around $\Delta_0$, the
Cooper-glass state. The excitation spectrum exhibits a broad gap
without quasiparticle peaks (Fig.\,\ref{spectre}), as observed in
tunneling experiments when coherence is broken, either at the
critical temperature
\cite{PRL_renner1998_T,Nat_Gomes2007,JPSJ_Sekine2016} or locally
within the vortex core \cite{PRL_renner1998_B}.

Below $T^*$ the CPG state is more favorable than the normal metallic
state but, due to pair-pair interactions, it is unstable to SC
condensation giving rise to the well-known critical dome
(Fig.\,\ref{phasediagram}, lower panel).

\vskip 0.2cm {\it The unconventional superconducting state} \vskip
0.2cm

The accurate temperature dependence of the quasiparticle spectra
measured by Sekine et al. \cite{JPSJ_Sekine2016} leads to the
counter-intuitive conclusion that the spectral gap value,
$\Delta_p$, remains constant right up to the critical temperature.
Significantly, this unique gap value is very robust as a function of
external perturbations, temperature, magnetic field and disorder.
Even though there are thermal pair excitations, so long as the
condensate is non vanishing, the gap value is pinned to $\Delta_p$.
Since the system remains statistically invariant in spite of the
perturbation, this leads to the novel concept of {\it ergodic
rigidity} specific to unconventional SC.

Therefore, we take as a principle that in the SC state, for which
$\beta^c\neq 0$, the gap value is imposed by the fundamental
properties of the system, the doping $p$ and the effective exchange
energy $J_{eff}$. The coherence in the SC phase imposes homogeneity,
which we obtain in the Voronoi scheme (equation\,\ref{equa_Jeff}) by
choosing all areas $\mathcal{A}_i$ to be equal. Thus, $\mathcal{A}_i
= \mathcal{A}_{\rm tot}/N_p$, where $N_p$ is the total number of
pairs and $\mathcal{A}_{\rm tot}$ the total area, guaranteeing
homogeneity. In addition, we assume that in the uniform SC state
$J_{eff} \rightarrow J$, where $J$ is the exchange energy. With
these considerations, we have\,:
\begin{equation}
\Delta_p = J\times (1-\frac{p}{p_c})
\end{equation}
as indicated in table\,\ref{table_param}. In the SC state the
spectral gap is thus uniquely determined by $p$ and $J$. Since in
the SC state the PPI is positive ($\beta^c
> 0$), comparing the gap equation\,(\ref{gap_equa}) with equation (\ref{delta0}) leads to
$\Delta_0 > \Delta_p$ and thus $J_{eff}> J$.

As a consequence of ergodic rigidity, the gap amplitude cannot be
the order parameter of the SC to PG transition. As opposed to
conventional SC, at finite temperature the system is not
characterized by quasiparticle excitations (pair breaking) but by
{\it pair excitations} following Bose-Einstein statistics
\cite{SciTech_Sacks2015}, leading to an unconventional shape of the
measured specific heat \cite{PhysicaC_Loram1994}.

Superconducting coherence is subtly hidden in the ground state
$\Delta_p$ but, contrary to BCS, it is not directly linked to the
superfluid density. Indeed, examining the gap
equation\,(\ref{gap_equa}), the SC order parameter is the PPI term,
$\beta^c(T) \propto N_{oc}(T)$, the latter being the number of
condensed pairs. It should be measurable in any experiment sensitive
to SC coherence such as Josephson effects
\cite{PhysLett_Josephson1962}. In quasiparticle tunneling, the
spectral gap remains constant, but the coherence peaks decrease
monotonically due to the increasing pairon excited states leading to
the pseudogap at $T_c$ \cite{EPJB_Sacks2016}.

The PPI energy is universally revealed in the shape of the
quasiparticle excitation spectrum beyond the gap energy measured by
tunneling and ARPES (see \cite{Revmod_Fisher2007} and refs.
therein). This well known `peak-dip-hump' fine structure is usually
interpreted in terms of a spin-collective mode
\cite{AdvPhys_Eschrig2006}. However, this compelling interpretation
fails to fit the shape of the experimental spectra and does not
account for the pseudogap phase.

The resolution of the microscopic Hamiltonian
\cite{SciTech_Sacks2015,SolStatCom_SMN2017} leads to the alternative
conclusion that, due to the PPI, the quasiparticles are now coupled
to excited pairon states (see Fig.\,\ref{quasiparticle}, right
panel). These `super-quasiparticles' are described by an unusual
dependence of the gap upon quasiparticle energy $E_k$\,:
\begin{equation}
\Delta_k(E_k)=\Delta_{0,k}-2\,\beta_k P_0(E_k)
\label{gap_equa_QP}
\end{equation}
where the anti-nodal direction is assumed. The corresponding DOS
(see Fig.\,\ref{spectre}) reveals the strong `peak-dip' structure in
the excitation spectrum seen in both cuprates
\cite{Revmod_Fisher2007} and iron-based superconductors
\cite{PRL_Chi2012,SciRep_Nag2016}.
Using the parameters of Table\,\ref{table_param}, the theoretical
curves fit accurately the experimental spectra for both classes of
materials\cite{SolStatCom_SMN2017}.
The peak-dip structure present at low temperature is thus a clear
signature of long-range SC coherence, while the mere presence of a
spectral gap is not.

We thus see clear deviations in the unconventional case from BCS in
all aspects, be it the SC ground state, the pair (boson) excitations
or the quasiparticle (fermion) excitations.

\vskip 0.2cm
{\it Universal dependence of the parameters on $J$}

\vskip 0.2cm

\begin{figure}[h!]
\centering \vbox to 6.5 cm{
\includegraphics[width=8.8 cm]{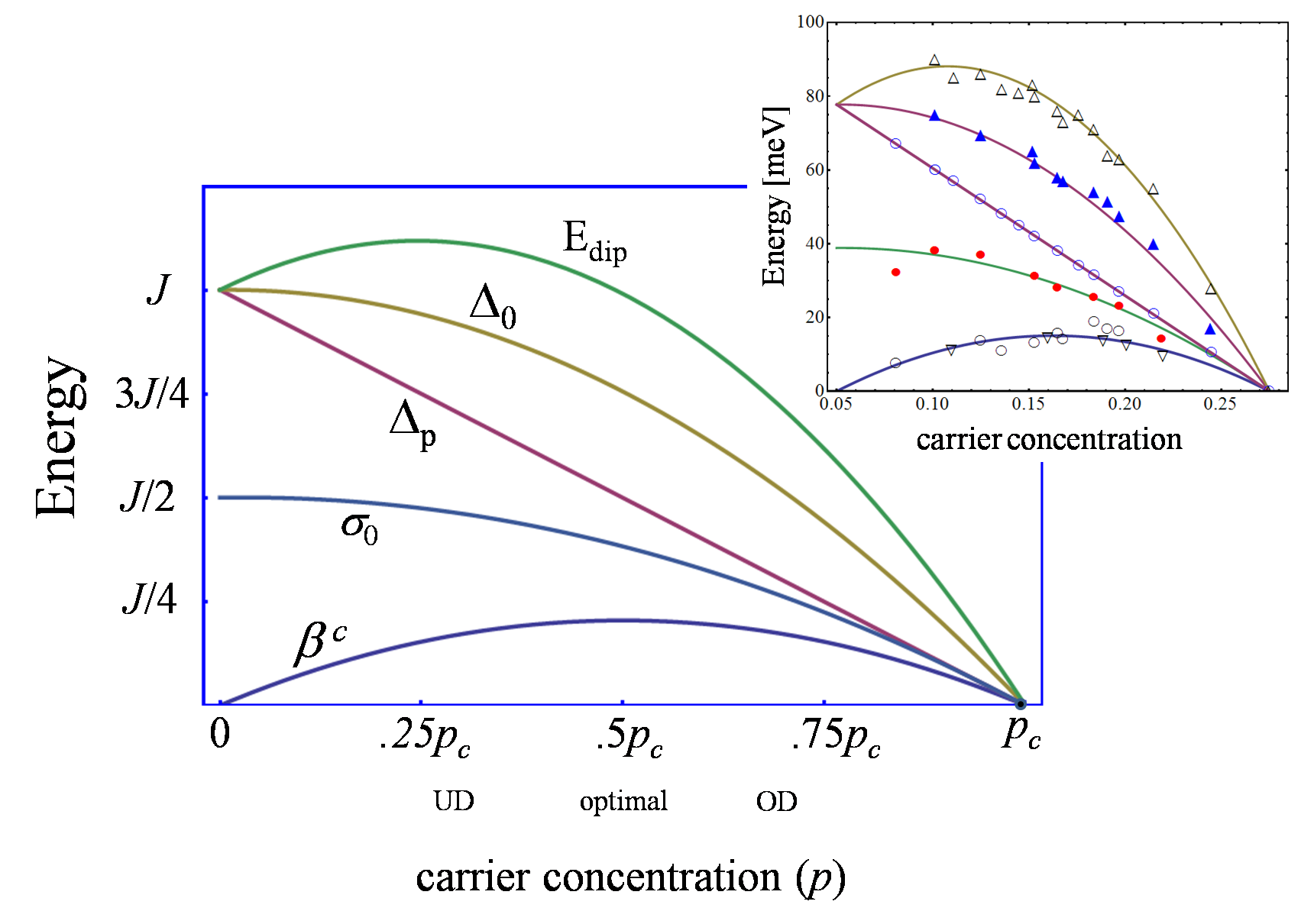}}
\caption{Universal doping dependence of the parameters (see Table
\ref{table_param})\,: The gap magnitude $\Delta_p$, the pair-pair
interaction $\beta$,  the excited state parameters $\Delta_0$ and
$\sigma_0$, and the dip position $E_{dip}$. The variation is shown
from $p=0$ at the SC to $p_c$ the critical doping. Note the
remarkable values\,: At optimal doping, $\beta^c$ is maximum while
the dip position is at $2\,\Delta_p$ such that $E_{dip} = J$. The
maximum value of $E_{dip}$ is at $p_c/4$. Inset: detailed fit for
Bi$_2$Sr$_2$CaCu$_2$O$_{8+\delta}$ using the equations of
Table\,(\ref{table_param}).}\label{param}
\end{figure}

How does the pair-pair interaction depend on the fundamental
parameters\,? In the mean-field treatment of the microscopic
Hamiltonian leading to equation (\ref{gap_equa}) the PPI is
repulsive and can be expressed as\,: $\beta_c \propto p\times
\Delta_p(p)$. From the detailed fits to the phase diagram
(Fig.\,\ref{param}, inset) the proportionality constant is found\,:
\begin{equation}\label{eq_beta}
\beta_c = 0,83\times J\left(\frac{p}{p_c}\right)\left(1-
\frac{p}{p_c}\right)
\end{equation}
where both $J$ and $p$ are measured from the SC onset.

This form of the mutual pair interaction can be visualized as the
lowering of the potential well due to the interaction with
neighboring pairs. Most significantly, both the gap magnitude
$\Delta_p$ and the interaction energy $\beta_c$ depend on a single
characteristic energy $J$. This shows unambiguously that the pair
formation (pairon) as well as the PPI depend on the same microscopic
mechanism, the intimate cooperation of holes with the underlying
antiferromagnetism.

As a function of the carrier concentration $p$, both $\Delta_p$ and
$\beta^c$ display a {\it universal dependence} upon doping
(Fig.\,\ref{param}). The binding energy $\Delta_p$ is given by the
exchange energy $J$ at low doping and decreases linearly with $p$,
while the PPI exhibits the universal dome, equation (\ref{eq_beta}),
as expected for the SC order parameter.

As shown in Fig.\,\ref{param}, all the parameters are simply
proportional to the exchange interaction energy, with typical value
$J\sim 100-130\,meV$ at half-filling in this class of materials
\cite{PRB_Sulewski1990}. Since the critical temperature scales with
$\beta^c$ ($\beta^c \simeq 2\,k_B\,T_c$ for Bi$_2$Sr$_2$CaCu$_2$O$_{8+\delta}$) this energy scale
explains the high critical temperature in cuprates.

Examining the gap equation (\ref{gap_equa}) we note that $\Delta_0$,
the average gap in the CPG state, is approximately given by\,:
$\Delta_0 \simeq \Delta_p + 1.2\,\beta^c$ leading to the effective
exchange energy $J_{eff}(p) \simeq J(1+1.2 p/p_c)$. Thus as function
of $p$, $\Delta_0(p)$ is a monotonic concave curve for the whole
doping range. In addition, as indicated in Table\,\ref{table_param},
the distribution width $2\sigma_0$ is very close to $\Delta_0$,
which is verified empirically (see inset in Fig.\,\ref{param}). Note
that $\Delta_0(p) - \Delta_p(p) \simeq 1.2\,\beta^c(p)$ defines the
{\it condensation energy} for any value of $p$.

As mentioned previously, the PPI is revealed in the dip position
$E_{dip}$ in the quasiparticle spectrum  which has a doping
dependence also shown in Fig\,\ref{param}. It lies above
$\Delta_0(p)$ for the whole doping range and is such that $E_{dip}
\simeq \Delta_0(p) + 1.2\,\beta^c$. The latter relation agrees very
well with experiment \cite{SciTech_Sacks2015}.

Many authors invoke the quasiparticle strong coupling to the spin
collective mode \cite{AdvPhys_Eschrig2006} to account for the dip
energy (relative to the gap $\Delta_p$) which is found to be $\sim
4.9\,k_B\,T_c$. In our work, the PPI is $\beta^c \simeq
1.8\,k_B\,T_c$ which gives $E_{dip} - \Delta_p$ to be
$4.4\,k_B\,T_c$, in good agreement. As mentioned, our interpretation
is completely different\,: The dip position is a direct consequence
of the static quasiparticle-pair interaction expressed in
Eq.\,(\ref{gap_equa_QP}).

The apparent contradiction can be lifted by noting that both depend
on the exchange energy $J$. In particular, at optimal doping our
model indicates the precise value $E_{dip} = J$. Since both
phenomena, the superconducting state and the paramagnon resonance,
depend on the same energy parameter they are indeed highly
correlated. On the other hand, as is well known in diverse fields,
even a strong correlation does not imply causality.

The interpretation of the phase diagram is thus as follows\,: the
subtle interplay between hole doping and antiferromagnetism leads to
a pairing, without boson exchange. Contrary to the BCS case where
Cooper pairs arise from the metallic state \cite{PR_Cooper1956},
here pairs emerge from an AF state. A crucial point is that the hole
pair state (pairon) preserves at best the symmetry of the AF
lattice. Such a pairing mechanism would not exist in a frustrated
spin system, such as in a triangular lattice, since in this case
there is no energy gain when forming hole pairs. This prediction is
in agreement with the absence of a SC transition in a Kagome-lattice
spin liquid \cite{PRX_Kelly2016}.

The pairons in this scenario cannot condense without the PPI and
would remain in the Cooper glass state. However their mutual
interaction, itself mediated by the same underlying AF local order,
allows for the condensation onto the homogenous and unique state
with energy $\Delta_p$. As a consequence, the mutual interaction
energy, $\beta^c$, as well as the pair formation mechanism, depend
on the unique energy $J$.

How does this picture translate to the quantum mechanical
formulation\,? Each hole pair in the AF sea, such as in
Fig.\,\ref{spins}, is surrounded by equivalent degenerate pair
positions where it can easily tunnel with no effect on the
surrounding AF symmetry. Its wave function is thus\,:
\begin{equation}\label{pairon}
\left|\Psi_{pairon}\right\rangle=\sum_{i,j}\,\alpha_{i,j}\,\hat
c_{i\downarrow}\hat c_{j\uparrow}\,\left|AF\right\rangle
\end{equation}
where $i$ and $j$ are nearest-neighbours and the Fermi operator
$\hat c_{i\downarrow}$ creates a hole in the AF half-filled state $\left|AF\right\rangle$ at
the site $i$. These pairon states are highly favorable and, when
interacting with neighboring pairons, i.e. the PPI process, allows
for a macroscopic SC wave function to be established.

\vskip 0.2cm {\it Conclusion} \vskip 0.2cm

In summary, in this article we show that the microscopic origin of
high-$T_c$ superconductivity is a subtle cooperation of holes with
the underlying local antiferromagnetism. First, preformed hole
pairs, the pairons, arise without `glue' in the AF background with a
binding energy directly related to the exchange AF energy $J$. The
superconducting phase emerges from an incoherent disordered pair
state, a Cooper-pair glass. As a result of pair-pair interactions,
they condense following the Bose-Einstein statistics onto a single
pair state. The pair binding energy $\Delta_p$ is anchored to a
value determined by the basic properties of the system, the exchange
energy $J$ and the doping $p$. The latter is remarkably robust with
respect to a perturbation, be it temperature or magnetic field,
suggesting the novel concept of {\it ergodic rigidity} at the heart
of the transition.

Contrary to BCS, the order parameter is the pair-pair interaction
energy $\beta^c$. The unconventional SC state is the subtle
compromise between pairon formation, of high binding energy, and
their repulsive interaction, both determined by the unique energy
$J$. The high-$T_c$ of cuprates thus follows. The single parameter
$J$ accounts quantitatively for the experimental phase diagram in
($T, p$) plane, the quasiparticle density of states and its
evolution as a function of temperature in both the superconducting
and the pseudogap phases.


\end{document}